\documentclass[showpacs,twocolumn,prb]{revtex4}

\usepackage{graphicx}
\usepackage{dcolumn}
\usepackage{amsmath}

\begin{document}

\preprint{HEP/123-qed}

\title{Terahertz magneto-spectroscopy of transient plasmas in semiconductors}

\author{M. A. Zudov$^1$}
\thanks{Present address: Department of Physics, University of Utah,
Salt Lake City, Utah 84112.}
\author{A. P. Mitchell$^2$}
\thanks{Present address: Lineup Technologies, Inc., Los Angeles,
California 90064.}
\author{A. H. Chin$^2$}
\thanks{Present address: UltraPhotonics, Fremont, California 94538.}
\author{J. Kono$^1$}
\thanks{To whom correspondence should be addressed.
http://www.ece.rice.edu/$\sim$kono. Electronic address: kono@rice.edu}

\address{
$^1$Department of Electrical and Computer Engineering,\\
Rice Quantum Institute and Center for Nanoscale Science and Technology,\\
Rice University, Houston, Texas 77005\\
$^2$W. W. Hansen Experimental Physics Laboratory, Stanford University,
Stanford, California 94305
}

\received{\today}

\begin{abstract}
Using synchronized near-infrared (NIR) and terahertz (THz) lasers, we have performed picosecond time-resolved THz spectroscopy of transient carriers in semiconductors.
Specifically, we measured the temporal evolution of THz transmission and reflectivity after NIR excitation.
We systematically investigated transient carrier relaxation in GaAs and InSb with varying NIR intensities and magnetic fields.
Using this information, we were able to determine the evolution of the THz absorption to study the dynamics of photocreated carriers.
We developed a theory based on a Drude conductivity with {\em time-dependent} density and {\em density-dependent} scattering lifetime, which successfully reproduced the observed plasma dynamics.
Detailed comparison between experimental and theoretical results revealed a linear dependence of the scattering frequency on density, which suggests that electron-electron scattering is the dominant scattering mechanism for determining the scattering time.
In InSb, plasma dynamics was dramatically modified by the application of a magnetic field, showing rich magneto-reflection spectra, while GaAs did not show any significant magnetic field dependence.
We attribute this to the small effective masses of the carriers in InSb compared to GaAs, which made the plasma, cyclotron, and photon energies all comparable in the density, magnetic field, and wavelength ranges of the current study.
\end{abstract}

\maketitle

\section{INTRODUCTION}
The advent of long-wavelength coherent sources, such as free-electron lasers (FELs),\cite{fel} parametric generators with difference frequency mixing,\cite{pg} terahertz (THz) antennas,\cite{THz} and quantum cascade lasers\cite{qcl} has brought an entirely new class of opportunities to study low-energy phenomena in solid state systems in the time domain and/or high-intensity regimes.
In particular, far-infrared (FIR) / THz pulses can directly probe low-energy dynamics in bulk and quantum-confined semiconductors, e.g., cyclotron resonance (CR),\cite{nurmikko,nicholas,singh,murdin,kono,mitchell}internal transitions of shallow donors\cite{seon,cole} and excitons,\cite{cerne,kono1,kent} phonons,\cite{planken,planken1} and intersubband transitions.\cite{elsaesser,elsaesser1,charlie,phillips,su}
In addition, small photon energies enhance the ponderomotive potential energy\cite{kent,chin} while precluding interband absorption and sample damage, leading to the possibility of {\em extreme} nonlinear optical behavior in semiconductors.\cite{chin,chin1}

In this paper we describe results of our study of the THz properties of photogenerated transient plasmas in semiconductors using a synchronized short-pulse THz $-$ near-infrared (NIR) laser system with picosecond time resolution, both in the absence and presence of an external magnetic field.
By simultaneously monitoring the temporal evolution of the transmission and reflection of a THz probe pulse after NIR excitation, we carried out a dynamical study of the Drude conductivity of transient plasmas.
More specifically, we were able to directly determine the density and scattering lifetime of photocreated transient carriers as functions of time, i.e., $n(t)$ and $\tau(t)$.

Another unique aspect of this technique lies in the fact that intraband FIR/THz spectroscopy is independent of whether the states involved are interband-active or not, thus providing a rare opportunity to directly probe nonradiative (or "dark") states.
Dynamics involving such states are not observable with conventional interband transient spectroscopies, e.g., time-resolved photoluminescence spectroscopy.
This unique ability makes it a powerful tool for providing insight into how optically created nonequilibrium electron-hole pairs lose their excess energies while relaxing toward the band edge through various scattering and thermalization processes before eventually recombining to luminesce.\cite{shah}

In our previous work,\cite{kono,mitchell} we described the first demonstration of picosecond time-resolved cyclotron resonance (TRCR) of photogenerated transient carriers by monitoring THz absorption as a function of magnetic field at fixed time delays between the NIR pump pulse and THz probe pulse.
In the present paper, we systematically investigated carrier relaxation in InSb and GaAs at different NIR intensities and magnetic fields.
Our calculations based on a Drude conductivity with $n(t)$ and $\tau(t)$ successfully reproduced the main observed features.
By fitting the theoretical reflectivity versus time to the experimental curves, we found a linear dependence of the scattering rate, $1/\tau$, on the carrier density, $n$.
This suggests that electron-electron scattering\cite{ridley} is the main factor in determining the carrier scattering lifetime inside the plasma in our density and time delay ranges.
The application of a magnetic field in InSb resulted in dramatic modifications in plasma dynamics whereas GaAs did not show any strong magnetic field dependence up to 8 Tesla.
This behavior of InSb can be attributed to the small effective masses of its carriers, which made the plasma, cyclotron, and photon energies all comparable, under our experimental conditions, and their subtle interplay led to the observed rich plasma dynamics.
Our theoretical simulations are in good qualitative agreement with the observations, supporting this explanation.

\section{EXPERIMENTAL METHODS}

\begin{figure}[tbp]
\includegraphics{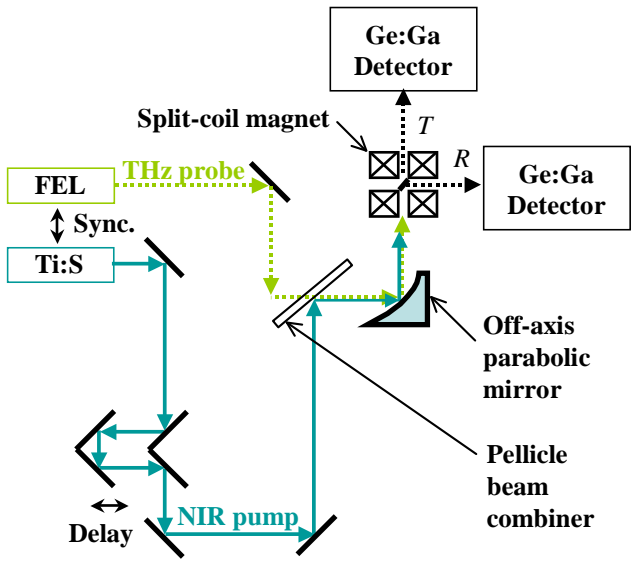}
\caption{Schematic diagram of the experimental setup used for
time-resolved two-color (NIR-THz) spectroscopy of transient plasmas in
semiconductors.}
\label{setup}
\end{figure}

The NIR laser source was a Ti:Sapphire laser seeding a regenerative amplifier.
The amplifier produced intense NIR ($\lambda_{{\rm NIR}} \approx$ 800 nm) pulses with pulse duration of $\sim$ 200 fs and pulse energies as high as $\sim$ 1 mJ at variable repetition rates up to 1 kHz.
The Stanford FEL\cite{schwettman} produced pulses (in macropulses, as described below) with wavelengths continuously tunable in the midinfrared (MIR) (3$-$15 $\mu$m) and FIR/THz (15$-$80 $\mu$m) with pulse durations ranging from 0.6 to 2 ps\cite{zudov} and pulse energies as high as $\sim$ 1 $\mu$J.
In the present study, the FEL wavelength was mostly fixed to $\lambda_{{\rm THz}}$ = 42 $\mu$m (or $\nu_{{\rm THz}}$ = 7.1 $\times$ 10$^{12}$ Hz or $\hbar \omega_{{\rm THz}}$ = 29.5 meV).

A schematic diagram of our experimental setup for the two-color spectroscopy experiments is illustrated in Fig. \ref{setup}.
The NIR output of the Ti:Sapphire system was directed through a computer controlled variable delay stage, after which it was spatially overlapped with the THz beam from the FEL using a Pellicle plate.
The two beams were thus made collinear as they were focused onto the sample using a parabolic mirror.
The NIR pulse excited nonequilibrium carriers across the band gap of the sample, which then absorbed a fraction of the incident THz probe pulse.
The transmitted and reflected THz beams were then recollimated and directed to liquid $^4$He-cooled Ge:Ga photoconductive detectors.
The THz output of the FEL was a pulse train of 10 Hz "macropulses" 5 ms in duration.
These macropulses each contained many ($\sim$ 60,000) $\sim$ 1 ps duration "micropulses" separated by 84.6 ns, corresponding to a repetition rate of 11.8 MHz.
The micropulse-to-micropulse energy fluctuations were factored out using a THz reference detector (not shown in Fig. \ref{setup}) before the sample.
The Ti:Sapphire oscillator was locked to the seventh harmonic of this repetition rate, i.e., 82.6 MHz.
Our synchronization electronics allowed us to select a single NIR pulse per FEL macropulse.
A combination of the optical delay stage and electronic delays in the synchronization allowed for selective delays from 0 to 84.6 ns with a picosecond resolution.
With this arrangement, we were able to compare the intensities of the transmitted and reflected THz pulses before and after the NIR pump pulse.
The amounts of photoinduced change in THz transmission and reflection were recorded as functions of time delay.  We then defined photoinduced absorption as
\begin{equation}
\Delta A = \frac {1 - R - T}{1 - R} - \frac {1 - R_0 - T_0}{1 - R_0}
\label{abs}
\end{equation}
where $R$ and $T$ are the reflectivity and transmissivity, respectively,
which are functions of time delay, and $R_0$ and $T_0$ are their
equilibrium values, i.e., before the arrival of the NIR pump pulse.
For InSb, we determined, by Fourier transform infrared spectroscopy,
that $R_0$ = 18\% (cf. theoretical value of $\sim 20$ \%) and
$T_0$ = 9\% for 42 $\mu$m radiation.

The InSb sample was undoped and had an electron density of 8.0 $\times$ 10$^{13}$ cm$^{-3}$ and a mobility of 8.3 $\times$ 10$^5$ cm$^2$V$^{-1}$s$^{-1}$ at 78 K.
The GaAs sample was semi-insulating, with excess arsenic.
We wedged both samples by $\sim 3^\circ$ to avoid multiple-reflection interference effects, and polished down to $\sim 150$ $\mu$m, still much thicker than the absorption depths of both InSb and GaAs at 800 nm.
This has an impact on the carrier dynamics, as significant carrier diffusion into the sample occurs (as discussed below).
The sample was placed inside a 9 T/1.5 K horizontal-bore split-coil magnet system (Oxford Instruments Spectromag 4000) with Sapphire cold windows and polypropylene room temperature windows.
The sample was tilted 45$^\circ$ with respect to the magnetic field, $B$, which was parallel to both laser beams.

\section{EXPERIMENTAL RESULTS}
\subsection{Power Dependence}
\begin{figure}[tbp]
\includegraphics{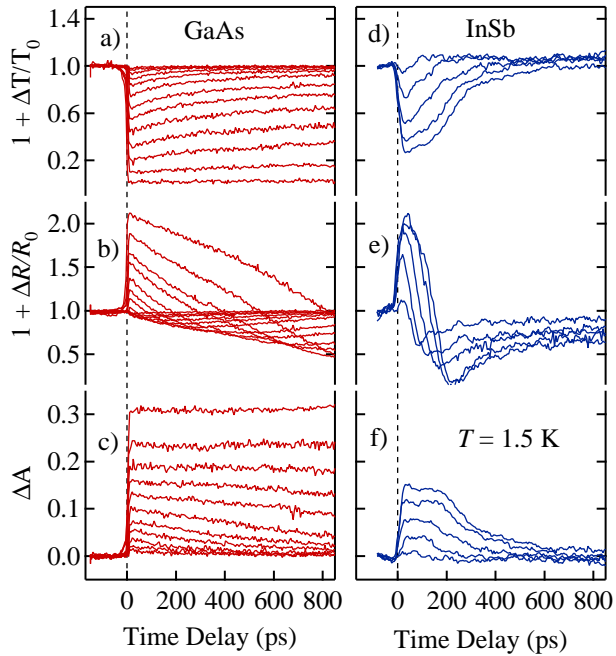}
\caption{Temporal evolution of the THz transmission, reflection and absorption
for different NIR pump intensities in the GaAs (left: a-c) and
InSb (right: d-f) samples.  The wavelength of the THz probe was
$\lambda_{{\rm THz}}$ = 42 $\mu$m (or $\nu_{{\rm THz}}$ = 7.1 $\times$
10$^{12}$ Hz or $\hbar \omega_{{\rm THz}}$ = 29.5 meV) and the sample temperature
was 1.5 K.}
\label{GaAsInSb}
\end{figure}
Typical zero-magnetic-field data for GaAs (left: a-c) and InSb (right: d-f)
are shown in Fig. \ref{GaAsInSb}.
The transmission, reflection, and absorption of the THz probe beam are
plotted against time delay.
The wavelength of the THz probe was 42 $\mu$m ($\nu_{{\rm THz}}$ = 7.1
$\times$ 10$^{12}$ Hz, $\hbar \omega_{{\rm THz}}$ = 29.5 meV) and the sample
temperature was 1.5 K.  Each panel shows multiple traces corresponding
to different NIR intensities, with the maximum NIR fluence at the
sample estimated to be $\sim$ 4 mJ/cm$^2$.
In both the GaAs and InSb samples, the photogenerated carriers cause an
abrupt drop (rise) in the THz transmission (reflection) at timing zero.
For example, at the maximum NIR intensity, the transmission drop is
$\sim 100$\% (complete transmission quenching) in GaAs and $\sim$ 70\% in
InSb.  The photoinduced absorption curves shown in Figs. \ref{GaAsInSb}(c)
and \ref{GaAsInSb}(f) were obtained from the measured transmission and
reflection curves using Eqn. (\ref{abs}).

The subsequent recovery of these photoinduced abrupt changes depends
critically on the relaxation properties of the sample under study.
It is clear from Fig. \ref{GaAsInSb} that there are significant differences
between GaAs and InSb.
In general, the GaAs sample shows smooth and monotonic temporal evolution
throughout the entire time range presented here (0$-$800 ps), whereas the
InSb sample shows much more complicated behavior, exhibiting dynamic changes
within the first $\sim$ 400 ps.
The decay of the transmission (or absorption) change in InSb is far from
monotonic, clearly showing multiple componets at high NIR intensities.
Its reflection dynamics are even more intriguing, exhibiting a sign
change (positive to negative) at a certain time delay, which sensitively
depends on the NIR pump intensity.
As discussed in Section IV, we can explain these dramatic differences between the two systems in terms of the importance of Auger processes, well known non-radiative carrier recombination especially important in narrow gap semiconductors like InSb.\cite{ridley}
The much shorter absorption depth of InSb as compared to GaAs also favors the importance of Auger recombination.
The sensitivity to the {\em total} carrier density, not the density of
the interband-active (or radiative) carriers alone,
distinguishes the current spectroscopic technique from conventional
transient spectroscopies based on interband transitions.

\begin{figure}[tbp]
\includegraphics{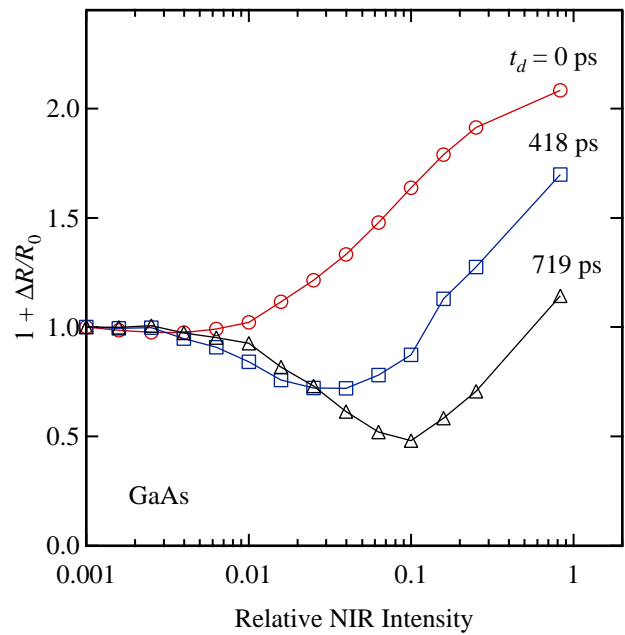}
\caption{Photoinduced reflectivity change as a function of NIR pump intensity for GaAs for fixed time delays of 0 ps, 418 ps, and 719 ps.
Lines connecting data points are guides to an eye.
It can be seen that with increasing NIR intensity the THz reflectivity can either increase or decrease, depending on the time delay.
We can also see, for time delays of 418 ps and 719 ps, that the reflectivity initially decreases and then increases monotonically.}
\label{GaAspower}
\end{figure}
The value of photoinduced reflectivity change is not a simple function of
time delay or NIR pump intensity.
Both its magnitude and sign depend on these parameters in a complicated
manner.
This is true even for GaAs, for which the reflection curves look smooth
and monotonic in Fig. \ref{GaAsInSb}(b).
To illustrate this point, we plotted the photoinduced reflectivity as a
function of NIR intensity for three different time delays in Fig.
\ref{GaAspower}.
Here the data were taken by varying the intensity of the NIR pump pulse
while the time delay was kept constant.
We can see that with increasing NIR intensity the THz reflectivity can
either increase or decrease, depending on the value of time delay.
Also, if the time delay is large enough, we see that the reflectivity
initially
decreases and then increases with the NIR intensity.
As we will see, the sign of the photoindiuced reflectivity change
is governed by the interplay between the plasma frequency,
$\omega_p$ ($\propto$ $\sqrt{n}$), and the THz photon frequency,
$\omega_{{\rm THz}}$.

\subsection{Magneto-Plasma Reflection}
\begin{figure}[tbp]
\includegraphics{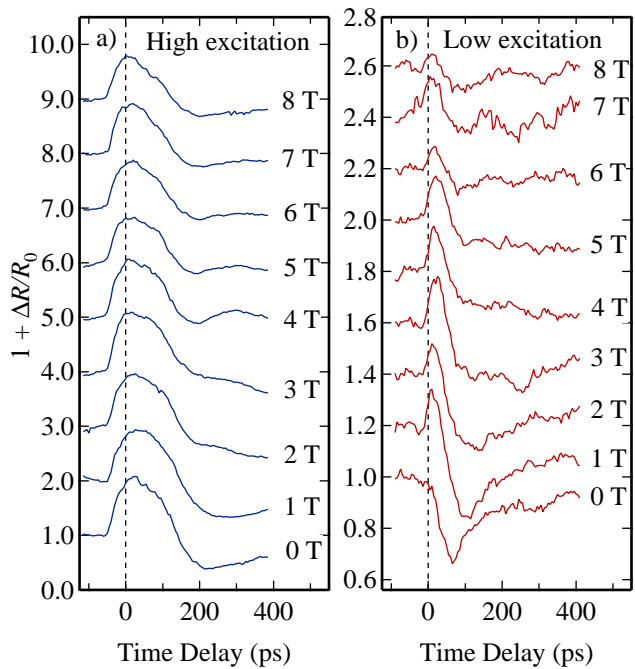}
\caption{Measured photo-induced reflection versus time delay at various fixed magnetic fields for (a) high ($\sim$4 mJ/cm$^2$) and (b) low ($\sim$15 $\mu$J/cm$^2$) NIR excitations in InSb at 1.5 K.
Traces are vertically offset for clarity.}
\label{mpr-exp}
\end{figure}

Since GaAs did not exhibit any magnetic field dependence in our accessible magnetic field range, here we concentrate on the data from InSb.
Figures \ref{mpr-exp}(a) and \ref{mpr-exp}(b) show photoinduced reflectivity versus time delay for InSb at various fixed magnetic fields for high ($\sim$ 4 mJ/cm$^2$) and low ($\sim$ 15 $\mu$J/cm$^2$) excitations, respectively, at a temperature of 1.5 K.
The data exhibit qualitatively different behavior under these different
excitation conditions.
All the traces in Fig. \ref{mpr-exp}(a) show significant photoinduced
{\em enhancement} in reflectivity immediately after photoexcitation.
This is because the plasma energy of the photocreated carriers initially
exceeds the photon energy of the THz probe (29.5 meV or 7.1 THz).
The estimated initial carrier density in the high-excitation case is of the
order of $\sim$ 10$^{19}$ cm$^{-3}$, which corresponds to a plasma
energy of $\sim$ 250 meV (or 60 THz).

The low-excitation data at $B$ = 0 T, on the other hand, shows a
reflectivity {\em drop} due to the created carriers, as shown in Fig.
\ref{mpr-exp}(b).  However, this completely opposite behavior very
quickly disappears as we increase the magnetic field from 0 to 1.5 T.
A small peak emerges in reflectivity, which grows in intensity with
increasing magnetic field, reaches a maximum at $\sim$ 3 T, stays roughly
constant up to $\sim$ 5 T, and finally goes away at higher magnetic field.
This very interesting behavior is explainable in terms of the tuning of
the plasma energy by the magnetic field,\cite{palik} or the interplay
between the plasma energy and the cyclotron energy, as discussed in
more detail in Section IV.B.

\section{DISCUSSION}
\subsection{Plasma Dynamics}
\label{theory}

We developed a theory for the reflectivity of a transient plasma as a
function of carrier density ($n$), scattering time ($\tau$), and magnetic
field ($B$).  After being created by the NIR laser beam, the carrier
population decreases on a picosecond time scale.
While there are a number of possible decay mechanisms, the dominant one,
especially in narrow gap semiconductors at high densities (which is the
case for InSb in this study), is known as
Auger relaxation.\cite{ridley,chazapis,ciesla,marchetti}
In this non-radiative decay mechanism, an electron and a hole recombine and
the resulting energy is transferred to a third carrier.  In this decay
mechanism, the carrier density decreases in a characteristic way:
\begin{equation}
\frac {dn}{dt}=-C_2 n^2,
\label{auger}
\end{equation}
where $C_2$ = 7.5 $\times$ 10$^{-9}$ cm$^3$/s is a reported Auger
coefficient for InSb.\cite{chazapis}
In our analyses we used a modified Auger coefficient $C_2^*$ = $\kappa C_2$
in order to account for other decay processes, e.g., carrier diffusion and
radiative recombination, which are not explicitly taken into account in
our model.  The density evolution $n(t)$ is then given by:
\begin{equation}
n(t)=\frac {1}{C_2 t + 1/n(0)},\,\, t\ge0
\label{dens}
\end{equation}
where $n(0)$ denotes the initial density of photocreated electron-hole
pairs.

The scattering rate $\tau^{-1}$ is taken to be a power law function of
the carrier density $n$, with coefficient $\alpha$, exponent $\beta$, and a small
fixed offset $\tau_i \approx 0.1$ ps used to account for density-independent
scattering mechanisms:
\begin{equation}
{\tau}^{-1}(t) =\alpha n^\beta(t)+ {\tau_i}^{-1}.
\label{sctime}
\end{equation}

The dielectric function of a semiconductor after ultrashort pulse excitation of a nonequilibrium plasma is given by:
\begin{equation}
\epsilon_\omega=\epsilon_{\infty}\left [1+\epsilon_{ph}(\omega) - \frac{\omega_p (t) ^2}{\omega (\omega - i/\tau (t) )}\right].
\label{df}
\end{equation}
Here $\epsilon_{\infty}$ = 15.68 is the dielectric constant for InSb,
$\omega_p$ = $[4\pi n(t) e^2/m_e\epsilon_{\infty}]^{1/2}$ is the plasma frequency
($m_e$ = 0.014$m_0$).
The phonon contribution $\epsilon_{ph}$ is calculated as:
\begin{equation}
\epsilon_{ph}(\omega)=\frac{\omega_L^2-\omega_T^2}{\omega_T^2-\omega^2+i\Gamma \omega}
\label{eph}
\end{equation}
where $\hbar \omega_L$ = 23.6 meV ($\hbar \omega_T$ = 22.2 meV) is the
energy of the longitudinal (transverse) optical phonon and $\Gamma$ =
0.35 meV.  For a fixed THz photon energy $\hbar \omega_{{\rm THz}}$ = 29.5 meV
(42 $\mu$m, 7.1 THz),
$\epsilon_{ph}$ $\approx$ $-$ 0.17 $-$ 0.005$i$ is the material parameter.

\begin{figure}[tbp]
\includegraphics{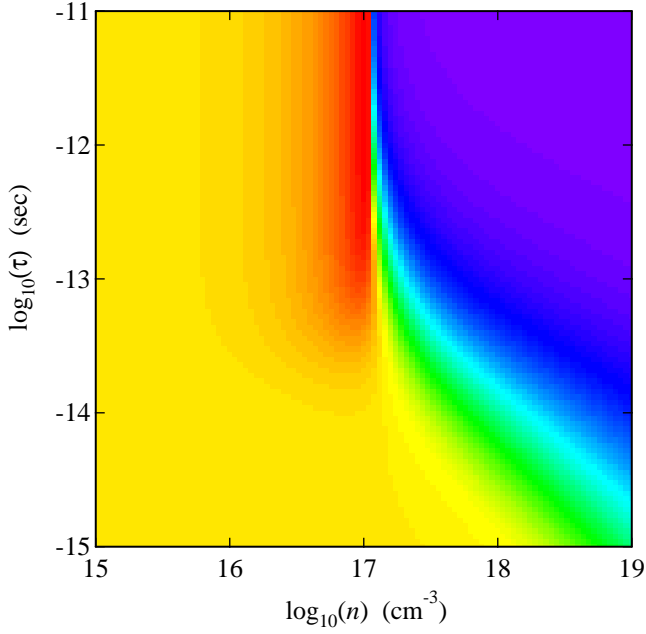}
\caption{Gray scale image of the reflectivity of a plasma created in
InSb as a function of carrier density $n$ and scattering time $\tau$
for a THz probe beam with a photon enenrgy of $\hbar \omega$ =
29.5 meV.}
\label{2d}
\end{figure}

At 45 degrees incidence, the reflection coefficient $r_\omega$ can be
calculated using a Fresnel formula:
\begin{equation}
r_\omega(t)=\frac{\epsilon_\omega \cos(\pi/4)-\sqrt{\epsilon_\omega-\sin^2(\pi/4)}}{\epsilon_\omega \cos(\pi/4)+\sqrt{\epsilon_\omega-\sin^2(\pi/4)}}.
\label{refl}
\end{equation}
In Fig. \ref{2d}, the reflectivity $R$ = $r_\omega r_\omega^*$ for
$\hbar \omega_{{\rm THz}}$ = 29.5 meV ($\nu_{{\rm THz}}$ = 7.1 THz)
is plotted as a function of scattering lifetime, $\tau$, and carrier
density, $n$, obtained through Eqns. (\ref{df}), (\ref{eph}), and (\ref{refl}),
in the ranges of $\tau$ = $10^{-15}-10^{-11}$ sec and
$n$ = $10^{15}-10^{19}$ cm$^{-3}$.
\begin{figure}[tbp]
\includegraphics{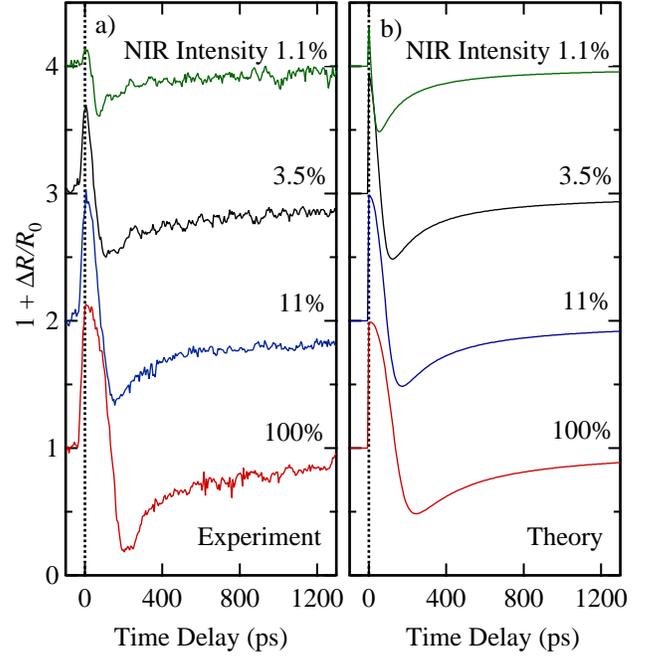}
\caption{Experimental (a) and calculated (b) photoinduced reflectivity of
InSb as a function of time delay for selected NIR pump intensities.
Traces are vertically offset for clarity.}
\label{fit}
\end{figure}
\begin{figure}[tbp]
\includegraphics{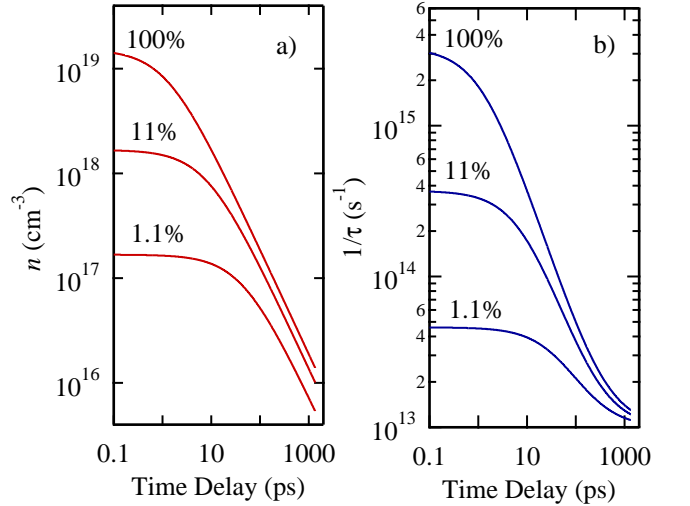}
\caption{Temporal evolution of the density (a) and scattering rate (b) in
InSb for selected NIR pump intensities.}
\label{nt}
\end{figure}
The temporal evolution of the normalized photoinduced reflectivity
due to transient plasmas is then computed as:
\begin{equation}
\frac {R(t)}{R_0}=1+\frac {\Delta R(t)}{R_0}= \frac{r_\omega(t)r_\omega^*(t)}{r_\omega(\infty)r_\omega^*(\infty)},
\label{pirefl}
\end{equation}
which can be directly compared with experimental data.  Here $R_0$ =
$r_\omega(\infty)r_\omega^*(\infty)\approx 0.198$ is the reflectivity in
equilibrium.

In Fig. \ref{fit} we present our experimental data (panel a) along with the fits (panel b) obtained using Eqn. (\ref{pirefl}).
We used the initial rise of the reflectivity, i.e., $\Delta R(0)/R_{0}$, to determine the initial carrier density, as well as the parameters $\alpha$ and $\beta$ in Eqn. (\ref{sctime}).
The initial density of photocreated carriers $n(0)$ scales linearly with the pump intensity, as expected.
We also found that most fits result in a value of $\beta$ very close to 1, suggesting the importance of electron-electron scattering.\cite{ridley}
The resulting expression for the scattering rate is as follows: $1/\tau$ $\approx$ 2.1 $\times$ 10$^{-4}$ $n$ [cm$^{-3}$] + 1/$\tau_i$ [cf. Eqn. (\ref{sctime})].
Using this expression for all NIR intensities, we then performed the time-evolution fits assuming the Auger-like decay of the carrier population [cf. Eqn. (\ref{dens})].
The multiplicative factor $\kappa$ showed a rather weak, but systematic, decrease from $\sim 17$ to $\sim 7$ with increasing NIR intensity.
The excellent agreement seen in Fig. \ref{fit} makes us believe that the carrier diffusion must be completed in less than $\sim$ 10 ps under our experimental conditions and its neglect in our model is justified.
In Fig. \ref{nt} we present the temporal evolution of the carrier density (a) and the scattering rate (b) calculated using Eqns. (\ref{dens}) and (\ref{sctime}) using the parameters derived from the fits.

\subsection{Magneto-plasma Dynamics}
\begin{figure}[tbp]
\includegraphics[scale = 0.55]{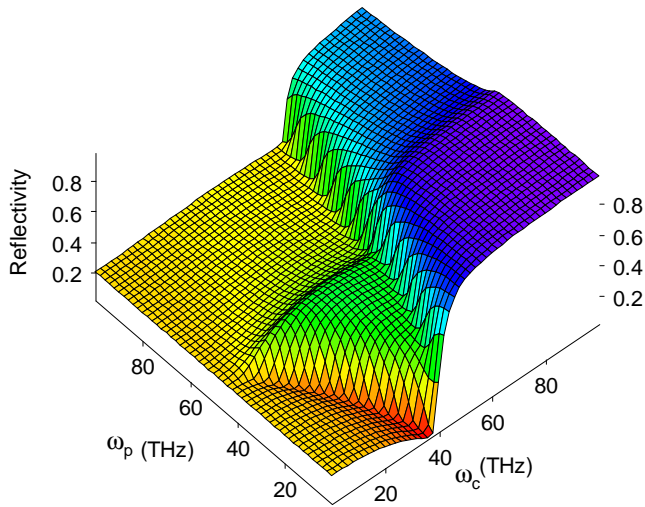}
\caption{Magneto-plasma reflectivity calculated by Eqn.
(\ref{reflectivityB}) for a THz beam with
$\hbar \omega$ = 29.5 meV as a function of plasma frequency ($\omega_p$)
and cyclotron frequency ($\omega_c$).
One can see that the reflectivity can be larger or smaller than the
equilibrium value in a complicated way, depending on the values of carrier
density and magnetic field.
The following parameters are used: $\omega\tau$ = 10.7, $\hbar \omega_L$
= 23.6 meV, $\hbar \omega_T$ = 22.2 meV, and $\Gamma$ = 0.35 meV.
[cf. Eqns. (\ref{eph}) and (\ref{dfb}).]}
\label{3d}
\end{figure}
\begin{figure}[tbp]
\includegraphics{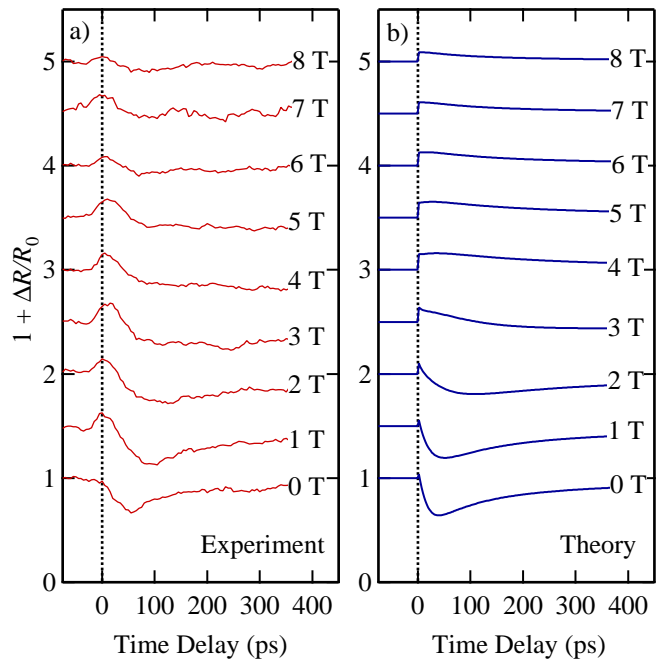}
\caption{Measured (a) and calculated (b) time-resolved photo-induced
reflection in InSb at a range of fixed magnetic fields.
Traces are vertically offset for clarity.}
\label{mpr-theory}
\end{figure}

When an external magnetic field is applied, Eqn. (\ref{df}) is modified, for
CR active ($+$) and inactive ($-$) polarizations, respectively, into:
\begin{equation}
\epsilon_{\omega,\pm}=\epsilon_{\infty}\left [1+\epsilon_{ph}(\omega) - \frac{\omega_p (t)^2}{\omega(\omega \pm \omega_c - i/\tau (t))}\right],
\label{dfb}
\end{equation}
where $\omega_c=eB/m_ec$ is the cyclotron frequency.
Correspondingly, Eqn. (\ref{refl}) changes to:
\begin{equation}
r_{\omega,\pm}=\frac{\epsilon_{\omega,\pm} \cos(\pi/4)-\sqrt{\epsilon_{\omega,\pm}-\sin^2(\pi/4)}}{\epsilon_{\omega,\pm} \cos(\pi/4)+\sqrt{\epsilon_{\omega,\pm}-\sin^2(\pi/4)}}.
\label{reflB}
\end{equation}
We then calculate the reflectivity, $R$, for a
linearly-polarized THz beam with frequency $\omega$ by taking an average
between the CR active and inactive circular polarization states:
\begin{equation}
R=\frac {r_{\omega,+}r_{\omega,+}^*+r_{\omega,-}r_{\omega,-}^*} {2}
\label{reflectivityB}
\end{equation}
This is plotted in Fig. \ref{3d} for a THz beam with $\hbar \omega$ = 29.5
meV as a function of $\omega_p$ and $\omega_c$.  Here the value of
$\omega\tau$ is fixed at 10.7, and $\hbar \omega_L$ = 23.6 meV,
$\hbar \omega_T$ = 22.2 meV, and $\Gamma$ = 0.35 meV are used for
Eqn. (\ref{eph}).  Figure \ref{3d} can be used for qualitatively predicting
THz magneto-reflection dynamics by treating the $\omega_p$ axis as the
time delay (since $\omega_p \propto \sqrt{n(t)}$) and the $\omega_c$ axis
as the applied magnetic field (since $\omega_c \propto B$).
Most importantly, one can clearly see from this figure that the instantaneous
value of reflectivity of a transient plasma can become larger or
smaller than the equilibrium value in a complicated manner, depending on the
instantaneous values of the carrier density and magnetic field.

The temporal evolution of the normalized photoinduced reflectivity for
a linearly-polarized THz probe with frequency $\omega$ is calculated as:
\begin{equation}
\frac {R(t)}{R_0} = 1+\frac {\Delta R}{R_0} = \frac{r_{\omega,+}(t)r_{\omega,+}^*(t)+r_{\omega,-}(t)r_{\omega,-}^*(t)}{r_{\omega,+}(\infty)r_{\omega,+}^*(\infty)+r_{\omega,-}(\infty)r_{\omega,-}^*(\infty)},
\label{pireflb}
\end{equation}
which can now be compared with experimental data.
Figure \ref{mpr-theory} illustrates (a) experimental and (b) theoretical
magneto-plasma reflection dynamics.
At zero magnetic field, the created plasma
{\em decreases} the reflectivity.  This behavior is due to the fact that
the plasma energy of the created carriers is lower than the THz photon
energy.  When this is the case, an increased carrier density translates
into a decreased refractive index, and hence, a decreased reflectivity.
Even a small applied magnetic field can dramatically change this behavior.
A small peak appears at zero delay, whose magnetude rapidly increases with
increasing magnetic field up to $\sim$ 3 T, saturates, and gradually decreases.
This intriguing behavior can be explained in terms of tuning of the plasma
edge by $B$, i.e., the interplay between $\omega_p$ and $\omega_c$, and is
successfully reproduced by our theoretical calculation
[Fig. \ref{mpr-theory}(b)].

\section{SUMMARY}

In summary, we performed picosecond two-color (NIR and THz) time-resolved
spectroscopy on GaAs and InSb.
We simultaneously monitored the dynamics of THz transmission and
reflection while we varied the strength of the applied magnetic field and
the time delay between the NIR and THz pulses.
We found that the photoinduced reflectivity dynamics are drastically
affected by the magnetic fields.
These results demonstrate the power and usefulness of this FIR technique
for investigating the dynamics of nonequilibrium carriers in semiconductors
at very low energy scales.

\section{ACKNOWLEDGMENTS}
We gratefully acknowledge support from NSF DMR-0049024, DMR-0134058 (CAREER),
DARPA MDA972-00-1-0034, AFOSR F49620-00-1-0349, the Japan Science and
Technology Corporation PRESTO Program, and the NEDO International Joint
Research Grant Program.
We thank H. Alan Schwettman, Todd I. Smith, Richard L. Swent, Takuji Kimura,
Eric Crosson, James Haydon, Jeffery Haydon, and George A. Marcus for their
technical support during this work at the Stanford Picosecond Free
Electron Laser Center and W. James Moore at the Naval Research Laboratory
for providing us with the high-quality InSb sample.

\bibliographystyle{unsrt}


\end{document}